\documentclass{pasj00}

\begin{document}
\SetRunningHead{K. Nakamura, T. Takiwaki, T. Kuroda, K. Kotake}{Systematic Features of 2D CCSNe}
\Received{2015/02/12}
\Accepted{2015/07/24}

\title{
Systematic Features of Axi-Symmetric Neutrino-Driven \\ Core-Collapse Supernova Models in Multiple Progenitors
}

\author{Ko \textsc{nakamura}}
\affil{Faculty of Science and Engineering, Waseda University, Ohkubo 3-4-1, Shinjuku, Tokyo 169-8555}
\email{nakamura.ko@heap.phys.waseda.ac.jp}

\author{Tomoya \textsc{takiwaki}
  \thanks{Present Address is 
  RIKEN, 
  2-1 Hirosawa, Wako, Saitama 351-0198
  }}
\affil{Center for Computational Astrophysics, National Astronomical Observatory of Japan, 2-21-1 Osawa, Mitaka, Tokyo 181-8588}

\author{Takami \textsc{kuroda}}
\affil{Department of Physics, University of Basel, Klingelbergstrasse 82, 4056 Basel, Switzerland}

\and
\author{Kei {\sc kotake}}
\affil{Department of applied physics, Fukuoka University, 8-19-1 Nanakuma Jonan, Fukuoka 814-0180}

%


\maketitle

\begin{abstract}
We present an overview of two-dimensional (2D) 
core-collapse supernova simulations employing neutrino transport 
scheme by the isotropic diffusion source approximation. 
We study 101 solar-metallicity, 
247 ultra metal-poor, and 30 zero-metal 
progenitors covering 
zero-age main sequence mass 
from $10.8 ~M_{\odot} $ to $75.0 ~M_{\odot} $. 
Using the 378 progenitors in total, 
we systematically 
investigate how the differences in the structures of these multiple 
progenitors impact the hydrodynamics evolution.
By following a long-term evolution over 1.0 s after bounce, 
most of the computed models exhibit neutrino-driven revival of the 
stalled bounce shock at $\sim$ 200 - 800 ms postbounce, 
leading to the possibility of explosion. 
Pushing the boundaries of expectations in previous one-dimensional (1D) 
studies, our results confirm that 
the compactness parameter $\xi$ that characterizes the structure 
of the progenitors is also a key in 2D to diagnose the properties of 
 neutrino-driven explosions.
Models with high $\xi$ 
undergo high ram pressure from the accreting matter onto the stalled shock,
 which affects the subsequent evolution of the shock expansion 
and the mass of the protoneutron star under the influence of 
 neutrino-driven convection and 
the standing accretion-shock instability. 
We show that the accretion luminosity becomes higher for models 
with high $\xi$, which makes the growth rate of the diagnostic 
explosion energy higher and the synthesized nickel mass bigger. 
We find that these explosion characteristics 
tend to show a monotonic increase as a function of the 
compactness parameter $\xi$. 
\end{abstract}

\section{Introduction}

The explodability of massive stars depends sensitively 
on the presupernova structures
 (e.g., \cite{Oconnor11,Ugliano12,couchott,Sukhbold14}). 
For low-mass progenitors with O-Ne-Mg core, the neutrino mechanism works 
successfully 
to explode in one-dimensional (1D) simulations because 
of the tenuous envelope \citep{kitaura}. 
For more massive progenitors 
with iron core, multi-dimensional (multi-D) effects such as 
neutrino-driven convection 
(e.g., \cite{bethe,herant,Burrows95,Janka96,mujan}) 
and the standing-accretion-shock-instability 
(SASI, \cite{Blondin03,Foglizzo06,Foglizzo07,Ohnishi06,Blondin07_nat,Iwakami08,Iwakami09,rodrigo09,jerome12,hanke12,thierry12,couch13,rodrigo14}, see
 \cite{thierry15} for a review) 
have been suggested to help the onset of the neutrino-driven explosion. 
Recently this has been confirmed by a number of self-consistent 
two-(2D) and three-dimensional (3D) simulations (e.g., \cite{Buras06a,Ott08,marek,bruenn13,suwa10,suwa14,BMuller12,BMuller13,Takiwaki12,taki14,Hanke13,dolence15,bruenn14,BMuller14}, see \cite{tony15,Janka12,burrows13,kotake12} 
for recent review)). 
Up to now, the number of these 
state-of-the-art models amounts to $\sim 40$
covering the zero-age main sequence (ZAMS) 
mass from $8.1 ~M_\odot$ (\cite{BMuller12}) 
to  $27 ~M_{\odot}$ (\cite{Hanke13}).

Based on stellar evolutionary calculations, on the other hand, 
 {\it hundreds} of CCSN progenitors are available now, depending 
 on a wide variety of the ZAMS mass, 
metallicity, rotation, and magnetic fields
(e.g., \cite{Nomoto88,WW95,Woosley02,Woosley07,Heger00a,Heger05,Limongi06}). 
These huge sets of CCSN progenitors, aided as well as 
by development of high-performance computers and numerical 
schemes, make {\it systematic} numerical study of CCSNe possible.

By performing general-relativistic (GR) 1.5D simulations for over 100 presupernova models using a leakage scheme, 
\citet{Oconnor11} 
were the first to 
point out that the postbounce dynamics and the 
progenitor-remnant connections are predictable basically
by a single parameter, compactness of the stellar core at bounce 
(see also \cite{oconnor13}).
Along this line, \citet{Ugliano12} 
performed 1D hydrodynamic simulations 
for 101 progenitors of \citet{Woosley02}.
By replacing the proto-neutron star (PNS) interior with an inner 
boundary condition, they followed an unprecedentedly 
long-term evolution over hours to days after bounce in spherical symmetry. 
Their results also lent support to the finding by \citet{Oconnor11} 
that the 
compactness parameter is a good measure to diagnose
 the progenitor-explosion and 
 the progenitor-remnant correlation (see also 
\cite{pejcha14,perego15}).
 
 Joining in these efforts but going beyond the previous 1D approach,
 we perform neutrino-radiation hydrodynamics simulations in two dimensions 
 using the whole presupernova series 
(101 solar-metallicity models, 247 ultra metal-poor models, 
and 30 zero-metal models) of \citet{Woosley02}.
Without the excision inside the PNS, we
 can self-consistently follow a long-term evolution 
 starting from the onset of core-collapse, bounce, 
neutrino-driven shock-revival, until the revived shock comes out of the iron core.
The goal of our 2D models is 
not to determine the very final fate of a massive star 
(which requires 3D-GR models 
 with detailed transport scheme (e.g., \cite{kuroda15})), 
but to study the systematic dependence of the 
progenitors' structure on the shock revival 
time, diagnostic explosion energy, 
mass of remnant object, and nucleosynthetic yields.
To this end, this study is the first attempt in the multi-D context.

Section \ref{sec-num} describes the numerical setup, 
including explanation of our numerical scheme (Section \ref{sec-scheme}), 
structure of the solar-metallicity 101 progenitors (Section \ref{sec-prog}), 
and discussion on effects of 
our choice of the outer boundary (Section \ref{sec-bc}). 
Results start from Section \ref{sec-res} where we first focus on the 
hydrodynamics evolution of the 101 solar-metallicity progenitors, and 
then move on to analyze the results in terms of the compactness parameters 
(Section \ref{sec-xi}). 
Section \ref{sec-uz} presents results of the 247 ultra-metal-poor and the 
30 zero-metal progenitors. We summarize our results and discuss their implications in Section \ref{sec-con}.

\section{Numerical setup}\label{sec-num}
\subsection{Numerical scheme}\label{sec-scheme}
The employed numerical methods are based on those in \citet{taki14}. 
Our 2D models are computed on a spherical polar grid of 
384 non-equidistant radial zones from the center up to 5000 km. 
Our spatial grid has a finest mesh spacing $dr_{\rm min} = 0.5$ km at the center 
and $dr/r$ is better than 1.8 \% at $r > 100$ km.
Our hydrodynamic scheme requires two ghost cell layers 
just above the outer boundary. 
We fix the density and velocity in the ghost cells to be the values there
of the progenitor models (see section \ref{sec-bc} for the effects of 
the outer boundary). 
We set 128 equidistant angular zones covering $0\leq \theta \leq \pi$ 
so that the angular resolution is $1.4^\circ$. 
We employ the equation of state (EOS) by \citet{LS91} 
with a compressibility modulus of $K=220$ MeV (LS220). 
For the calculations presented here, self-gravity is computed by a Newtonian monopole approximation and our code is updated, from the ZEUS-MP \citep{Hayes06}
 code as was used in \citet{taki14},
 to use high-resolution shock capturing scheme with an approximate Riemann solver of \citet{Einfeldt88}.
As described in \citet{nakamura14}, 
we take into account explosive nucleosynthesis and 
the energy feedback into hydrodynamics 
by solving a 13 $\alpha$-nuclei network including 
$^4$He, $^{12}$C, $^{16}$O, $^{20}$Ne, $^{24}$Mg, $^{28}$Si, 
$^{32}$S, $^{36}$Ar, $^{40}$Ca, $^{44}$Ti, $^{48}$Cr, $^{52}$Fe, and $^{56}$Ni. 
The nuclear energy compensates for energy loss 
via endothermic decomposition of iron-like NSE nuclei to lighter elements
(see Appendix of \citet{nakamura14} 
for more details). 

To solve spectral transport of electron- ($\nu_e$) and anti-electron neutrinos 
(${\bar \nu}_e$), 
we employ the isotropic diffusion source approximation (IDSA, \cite{idsa}).
We take a ray-by-ray approach, in which the neutrino transport 
is solved along a given radial ray assuming that the 
hydrodynamic medium for the 
 direction is spherically symmetric (e.g., \cite{Buras06a}). 
Although one needs to deal with the lateral 
transport more appropriately (e.g., \cite{sumi15,dolence15}), 
this approximation
 is useful because of the high computational efficiency in parallelization,
 which allows us to explore the more systematic progenitor survey
 based on the radiation-hydrodynamics models than ever in this study. 
Regarding heavy-lepton neutrinos 
($\nu_x = \nu_{\mu}, \nu_{\tau}, \bar{\nu}_{\mu}, \bar{\nu}_{\tau}$), we employ
 a leakage scheme to include the $\nu_x$ cooling via pair, 
photo and plasma processes (see \cite{taki14} 
for more details). 
To induce non-spherical instability, we add initial 
seed perturbations by zone-to-zone random density variations with 
an amplitude less than 1\%.  

There is still a debate about 
whether explosions are obtained more easily 
in low-resolution simulations than in high-resolution ones and 
how much resolutions are needed to obtain a convergence (e.g., 
 \citet{radice}). 
In Appendix \ref{sec-app}, we discuss the resolution dependence of 
 our 2D results.

\subsection{Progenitor models}\label{sec-prog}
\begin{figure}
  \begin{center}
    \FigureFile(90mm,90mm){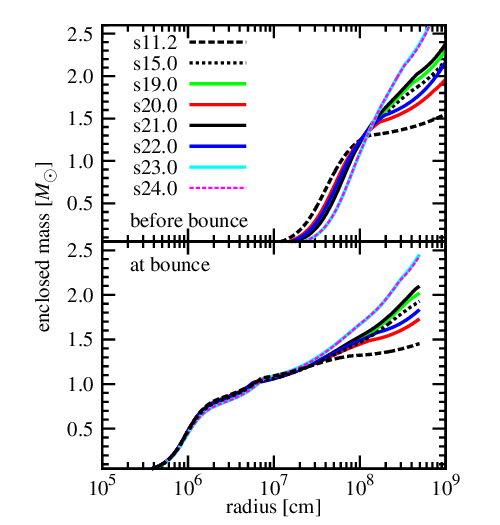}
  \end{center}
  \caption{
  Mass distribution of some selected models 
  at a pre-collapse stage (top panel) 
  and at the time of core bounce (bottom).
  }\label{fig-mr}
\end{figure}

The investigated solar-metallicity progenitors with iron cores 
\citep{Woosley02} are given 
in $0.2 ~M_{\odot}$ steps between $10.8 ~M_{\odot}$ (s10.8) 
and $28.2 ~M_{\odot}$ (s28.2), 
further from $29 ~M_{\odot}$ (s29) 
up to $40 ~M_{\odot}$ (s40) 
in $1.0 ~M_{\odot}$ steps, and a $75 ~M_{\odot}$ model (s75), 
101 progenitors in total. 
The structure of these stars, such as the density profiles and the pre-collapse 
mass distributions has been already described in \citet{Ugliano12}. 
Here for convenience, we show the mass distribution of some selected models 
at a pre-collapse stage and the time of core bounce (Figure \ref{fig-mr}). 
Before bounce, the distribution varies from models to models, 
especially at the outer radius. 
The mass distribution of the collapsing core dynamically changes after 
the onset of collapse, but at bounce, the mass distributions 
remain almost identical within $\sim 2 \times 10^7$ cm 
among the different progenitor models (bottom panel).

To characterize the progenitor structures, 
we estimate the compactness parameter $\xi$
following \citet{Oconnor11}
as the ratio of mass $M$ and the enclosed radius $R(M)$,
\begin{equation}
\xi_M  \equiv \frac{M/M_{\odot}}{R(M)/1000{\rm km}}.
\label{eq-xi}
\end{equation}
The previous studies used $M=2.5 \, ~M_{\odot}$ \citep{Oconnor11,Ugliano12} 
or $1.75 \, ~M_{\odot}$ \citep{oconnor13} 
and estimated $\xi_M$ 
at the time of core bounce. 
On the other hand, 
the outer radius of our computational domain (5000 km) is too small to 
contain $2.5 \, ~M_{\odot}$ for all models 
and even  $1.75 \, ~M_{\odot}$ for some less massive models 
(see Figure \ref{fig-mr}). 
In this paper, we estimate $\xi_M$ at $M=2.0$ and $2.5 \, ~M_{\odot}$ 
($\xi_M = \xi_{2.0}, \,\xi_{2.5}$) 
directly from the progenitor models.
It should be noted that 
our definition of $\xi_{2.5}$ gives almost the same value 
as the compactness estimated at bounce, 
because the radius $R$ enclosing $2.5 \, ~M_{\odot}$ is 
far from the center and the radial velocity $v_R$ there is very small 
(e.g., for s15.0 model, $R=1.7 \times 10^{9}$ cm and 
$v_R=-6.8 \times 10^6\,\,{\rm cm \,\, s}^{-1}$).
By comparing the top to bottom panel of Figure \ref{fig-mr}, 
the position of the outer envelope ($\gtrsim 10^{8}\,{\rm km}$) 
changes very slightly. This is because of the long dynamical time scale there 
compared to the short time period before bounce ($\sim 200$ ms).
Actually $\xi_{2.5}$ of s15.0 model in our definition is 0.149, 
which is very close to the value (0.150) estimated by \citet{Oconnor11}
at bounce. 

\begin{figure}
  \begin{center}
    \FigureFile(90mm,90mm){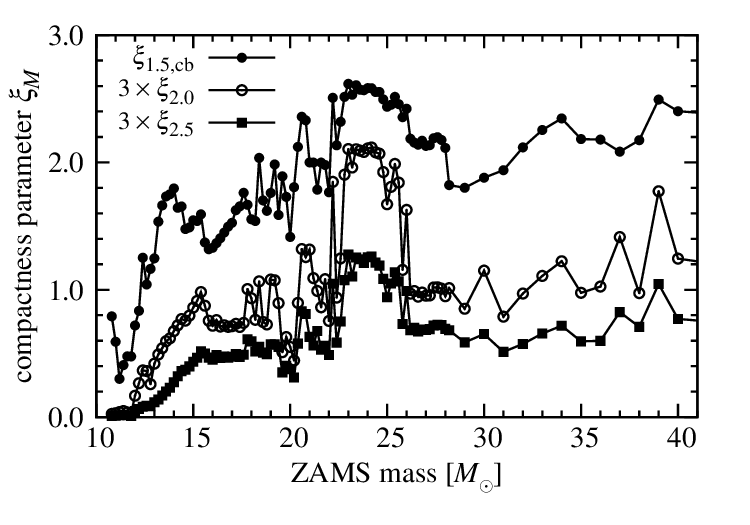}
  \end{center}
  \caption{
  Three choices of the compactness parameters 
  $\xi_M$ as a function of ZAMS mass. 
  They are estimated at $M=1.5 \, ~M_{\odot}$ ($\xi_{1.5,{\rm cb}}$), 
  $2.0 \, ~M_{\odot}$ ($\xi_{2.0}$), 
  and $2.5 \, ~M_{\odot}$  ($\xi_{2.5}$) from top to bottom.
  Note that $\xi_{1.5,{\rm cb}}$ is estimated at the time of core bounce 
  as in the previous studies. 
  The others are from pre-collapse data and enhanced by a factor of three 
  for comparison. 
  Model s75 is out of this plot 
  but has 
  $\xi_{1.5,{\rm cb}} = 2.0$, $\xi_{2.0} = 0.17$, and $\xi_{2.5} = 0.11$. 
  }\label{fig-xi}
\end{figure}

Figure \ref{fig-xi} compares $\xi_M$ estimated at 
$M=1.5 \, ~M_{\odot}$ ($\xi_{1.5,{\rm cb}}$), 
$2.0 \, ~M_{\odot}$ ($\xi_{2.0}$), 
and $2.5 \, ~M_{\odot}$  ($\xi_{2.5}$).
Note that $\xi_{1.5,{\rm cb}}$ is estimated at the time of core bounce, 
whereas the others are directly estimated from the progenitor data. 
All of these profiles show a similar trend, 
for example a high compactness bump
in the $22$ to $26 ~M_{\odot}$ mass range. 
Zigzag variations of the compact parameter can be seen
for all the lines with the different choice of $M$,
although the amplitude becomes smaller for $\xi_M$ with larger $M$. 
Among all the 101 models, 
the s11.2 progenitor has the smallest value of the compactness parameter, 
which is easily seen for the choice of $\xi_{1.5,{\rm cb}}$ (filled circles). 
The quantities of the compactness parameters for some representative models 
are listed in Table \ref{tbl-xi}. 
It includes $\xi_{1.5}$ from the pre-collapse data 
and $15 ~M_\odot$ progenitors from 
\citet{WW95} (WW15) and  \citet{Woosley07} (WH15) 
as a reference.
Metal-deficient progenitors are discussed in Section \ref{sec-uz}.

\begin{table}
  \caption{
  Variations of the compactness parameters 
  from the solar-metallicity progenitors \citep{Woosley02}}\label{tbl-xi}
  \begin{center}
    \begin{tabular}{lcccc}
      \hline 
      model & $\xi_{1.5,{\rm cb}}$ & $\xi_{1.5}$ & $\xi_{2.0}$ & $\xi_{2.5}$\\
      \hline
       s11.2 & 0.300 & 0.195 & 0.014 & 0.005 \\
	s15.0 & 1.546 & 0.862 & 0.298 & 0.149\\
	s19.0 & 1.761 & 0.911 & 0.374 & 0.194\\
	s20.0 & 1.414 & 0.671 & 0.187 & 0.127\\
	s21.0 & 2.000 & 0.971 & 0.455 & 0.215\\
	s22.0 & 1.766 & 0.868 & 0.258 & 0.165\\
	s23.0 & 2.617 & 1.000 & 0.720 & 0.434\\
	s24.0 & 2.583 & 0.998 & 0.721 & 0.427\\
	s30 & 1.880 & 0.938 & 0.394 & 0.222\\
	s40 & 2.402 & 0.990 & 0.427 & 0.263\\
	s75 & 2.004 & 0.890 & 0.168 & 0.112\\ 
	\hline
	WW15 & 0.765 & 0.592 & 0.194 & 0.085\\ 
	WH15 & 1.220 & 0.871 & 0.335 & 0.181\\ 
      \hline
    \end{tabular}
  \end{center}
\end{table}

\begin{figure}
  \begin{center}
    \FigureFile(90mm,90mm){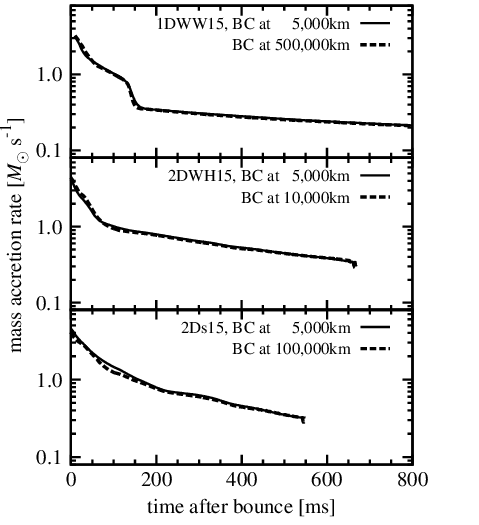}
  \end{center}
  \caption{
  Mass accretion rate of $15 ~M_\odot$ progenitor models 
  with outer boundaries at different radii. 
  The lines given by 
  2D simulations of WH15 model (middle panel), and s15 model (bottom), 
  are truncated when the shock touches 500 km 
  at $t_{\rm pb} \sim 670$ ms, and $\sim 550$ ms, respectively. 
  1D simulations of WW15 model (top) do not present shock revival.
  The difference of the boundary position does not affect the mass accretion rate 
  in all the chosen progenitors.
  }\label{fig-mdot}
\end{figure}

\subsection{Boundary condition}\label{sec-bc}
Our simulation domain is limited within the radius of 5,000 km 
so that we can reduce computational cost 
and carry out 2D self-consistent simulations for the 378 progenitors. 
This relatively small spatial domain (5,000 km), however, might affect the 
 hydrodynamics evolution long after bounce. 
To clarify this, we check the effects of the outer boundary by 
estimating the mass accretion rate at different radii. 
We focus on the
mass accretion rate because it predominantly affects 
the explosion properties as we will discuss later. 

In Figure \ref{fig-mdot}, we show the mass accretion history 
of the three progenitor models with the same ZAMS mass 
(15 $M_{\odot}$); 
WW15 \citep{WW95} (top panel), WH15 \citep{Woosley07} (middle panel)
, and s15 \citep{Woosley02} (bottom panel), respectively. 
Following \citet{BMuller14}, we estimate the mass accretion rate
 at the radius of 500 km in Figure \ref{fig-mdot}.
The radius of the outer boundary
 ($R_{\rm out}$) is taken either at 5000 km
 or at the more distant radius 500,000 km (top panel), 10,000 km 
(middle panel), or 100,000 km (bottom panel), respectively.
Note that the shock revival is not obtained in the WW15 model that is 
computed in 1D, whereas the shock revival is obtained at 
$t_{400} =688\,$-$\, 691$ ms 
and $556\,$-$\,571$ ms in the 2D simulations for the WH15 and s15 models,
 respectively. In the 2D models, the mass accretion rate 
is only shown before the shock revival (e.g., middle and bottom
 panels of Figure 3) because the mass accretion rate 
at 500 km can be affected by the non-radial motions of the expanding shock.

From Figure \ref{fig-mdot}, it is shown that the mass accretion rate (before
 the shock reaches at the radius of 500 km) is very similar for models 
with the different $R_{\rm out}$. 
It should be noted that the mass accretion rate of the WW15 model shows
a sudden drop from $0.8 ~M_\odot \, {\rm s}^{-1}$ to $0.3 ~M_\odot \, {\rm s}^{-1}$ 
at $t_{\rm pb} \sim 150$ ms ($t_{\rm pb}$; postbounce time) 
then gradually decreases to  $0.2 ~M_\odot \, {\rm s}^{-1}$. 
This is in a good agreement with the previous results 
using the same progenitor model (e.g., Figure 1 in \citet{Murphy08} 
and Figure 1 in \citet{hanke12}). 
All of these facts support that our boundary conditions 
well imitate the density structure out of the boundary 
of these three $15 ~M_\odot$ models 
and the boundary effect is not significant for these models.

\begin{figure}
\begin{center}
\FigureFile(90mm,90mm){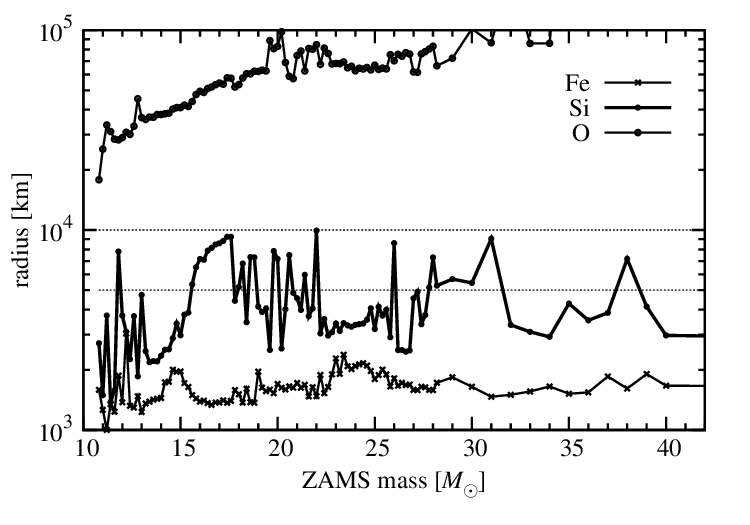}
\end{center}
\caption{
Radius of Fe core surface, Si/O interface, and top of O layer 
for solar-metallicity models.
}
\label{fig-rcom}
\end{figure}

\begin{figure*}
\begin{center}
\FigureFile(170mm,170mm){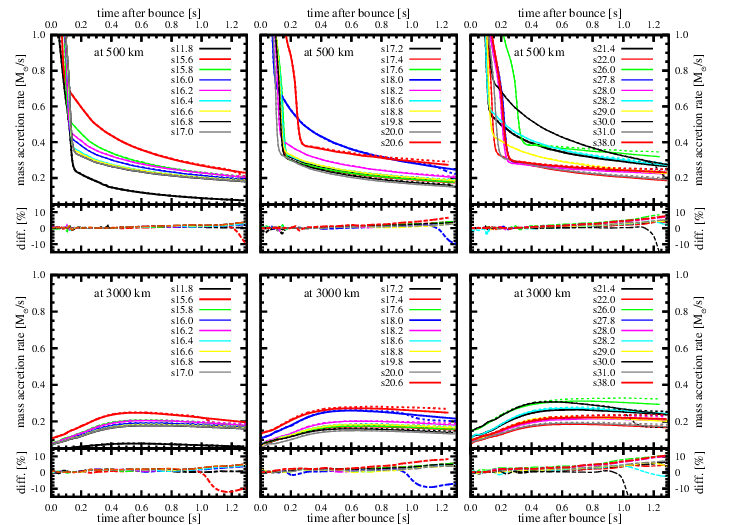}
\end{center}
\caption{
Comparison of the mass accretion rate between 1D models 
with the outer boundary at the radius of $r=$ 5,000 km ($\dot{M}_5$, solid lines) 
and 10,000 km ($\dot{M}_{10}$, dotted). 
Twenty-nine solar-metallicity models 
which have the Si/O interface between the radii of 5,000 km and 10,000 km 
are shown (see also Figure 4). 
Top: the mass accretion rate estimated at $r=500$ km, 
representing the accretion onto a stalling shock front. 
Bottom: same as the top panels 
but estimated at $r=3,000$ km, 
representing the accretion rate onto an expanding shock front in 2D models. 
Relative differences between the models 
with the two different outer boundary positions 
$(\dot{M}_{10} - \dot{M}_5)/\dot{M}_5$
are presented at the lower panel of each plot in
percentage. 
}
\label{fig-mdot2}
\end{figure*}

However, what if there would exist a large number of progenitors that 
would have shell interfaces 
with sharp density gradients beyond our computational domain (5000 km)?
Figure 4 shows the position of the shell interfaces for all the 
solar-metallicity models employed in our study. 
Here we define the radius of 
silicon (Si) layer at the position where the most abundant element changes 
from silicon to oxygen (O). Out of the 101 models, 
we identified 29 models that have the 
Si/O interface at the radius between 5,000 and 10,000 km 
(horizontal dotted lines). 
For these models having the shell interfaces above 5,000 km, 
we conduct 1D simulations and plot the mass-accretion rate history 
in Figure \ref{fig-mdot2} 
to examine the effects of the outer boundary positions.
In each model,  we vary the position of the outer boundary, either 
at the radius of 5,000 km (shown by solid lines) or at 
10,000 km (dotted lines). 

From each panel of Figure \ref{fig-mdot2}, 
it is shown that 
all of the examined models with the outer boundary at $r=5,000$ km 
show very small differences (less than a few percent) 
from the ones with the boundary at $r=10,000$ km before $t_{\rm pb} \sim 0.9$ s. 
In this epoch, the material initially located at $r > 5,000$ km is still 
at $r > 3,000$ km and the boundary effect is almost negligible.
After $t_{\rm pb} \sim 0.9$ s, the material initially out of the radius of 
5,000 km begins passing through the radius of $r=3,000$ km. 
Among the 29 progenitors (with shell interfaces above 5,000 km, 
see also Figure 4), 
only 3 models show a remarkable feature in the mass accretion rate. 
As shown in the bottom panels of Figure 5,  
s15.6 (a red dotted line in the left panel), 
s18.0 (a blue dotted line in the middle panel), 
and s30.0 (a black thin-dotted line in the right panel) 
with the outer boundary at $r=10,000$ km 
show a sudden drop ($\gtrsim 10 ~\%$) 
at $t_{\rm pb} \sim 1.0$ s, 
followed by the drop of the accretion rate 
in the top panels at $t_{\rm pb} \sim 1.2$ s. 
This decrease of the accretion rate 
in these three models 
is caused by a density jump at the Si/O interface 
initially located at the radius $r > 5,000$ km, 
which cannot be taken into account by the models 
with the outer boundary at $r=5,000$ km. 
In contrast to 
the models s15.6, s18.0, and s30.0, 
$10 ~\%$ level increases are observed 
in the models s20.6, s26.0, s28.0, and s38.0. 
This is because 
these models have a relatively high density envelope, 
for which our choice of the boundary position (at 5,000 km) 
underestimates the mass accretion rate compared to that at 10,000 km. 
There might be such models with high density envelope 
in the rest 72 progenitors 
other than the four of 29 progenitors discussed here. 

To quantify the boundary effects on these models, 
we perform 2D simulations for s38.0 model in the following way. 
Using the same setups as our fiducial model, 
the density out of the maximum shock radius is manually changed 
when the shock reaches 3000 km. 
We find that increase (decrease) of the density by $10 ~\%$ 
results in $2.8 ~\%$ ($0.53 ~ \%$) change of the diagnostic energy 
and negligibly small difference of the PNS mass  ($< 0.03 ~ \%$)
at the time when the shock reaches the outer boundary at 5000 km. 
This is simply because of 
a small mass accretion rate ($< 0.3 ~ M_\odot$ s$^{-1}$) 
at this phase 
and a short period until the simulations are terminated 
when the shock reaches the outer boundary. 
Thus, we conclude that 
the boundary effects on the mass accretion rate 
in the late postbounce phase 
is not influential to our systematic study.


\begin{figure}
\begin{center}
\FigureFile(70mm,70mm){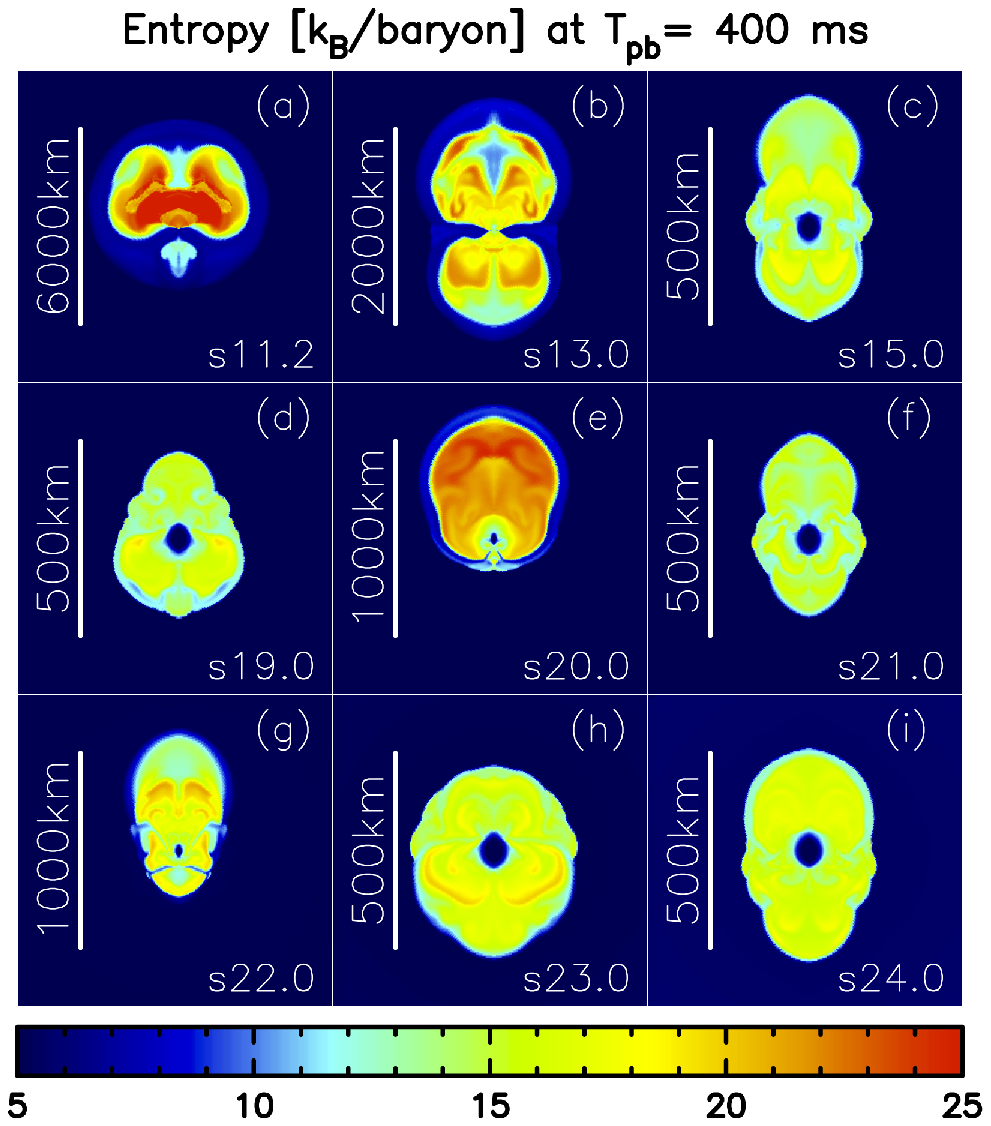}
\end{center}
\caption{Entropy distributions in unit of $k_{\rm B}$ per baryon 
for selected nine models at $t_{\rm pb}=400$ ms. 
Shown are models 
s11.2 (a)
to s24.0 (i), 
from top-left to bottom-right.
Note the different scale in each panel.}
\label{fig-entall1}
\end{figure}

\begin{figure}
\begin{center}
\FigureFile(70mm,70mm){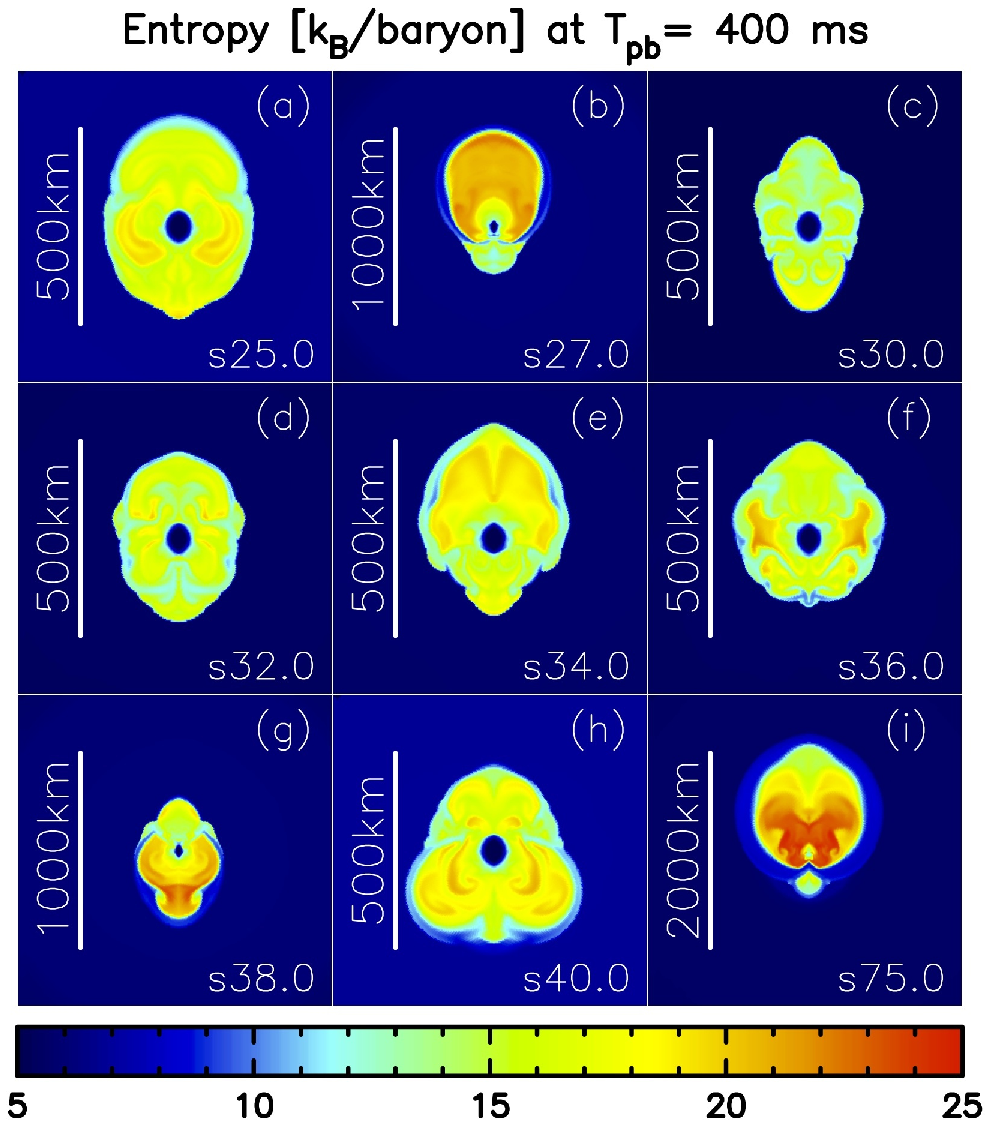}
\end{center}
\caption{Same as Figure \ref{fig-entall1} 
but for models 
s25.0 (a)
to s75.0 (i), 
from top-left to bottom-right.
}
\label{fig-entall2}
\end{figure}

\section{Results}\label{sec-res}
 For all the solar-metallicity 101 models 
 (and also the additional 277 models discussed in Section \ref{sec-uz}),
 the bounce shock stalls in a 
 spherically symmetric manner and only after that, we observe 
a clear diversity of the multi-D hydrodynamics evolution in
 the postbounce (pb) phase.
Figures \ref{fig-entall1} and \ref{fig-entall2} show a snapshot of entropy distribution 
for selected 18 solar-metallicity models at $t_{\rm pb}=400$ ms.
For some less massive progenitors 
(e.g., model s11.2 in Figure \ref{fig-entall1}(a)),
the shock is reaching close to the outer boundary of the computational
 domain with developing 
 pronounced unipolar and dipolar shock deformations. At this time, 
the shock of the most massive progenitor 
(s75.0 in Figure \ref{fig-entall2}(i)) 
is reaching an average radius of $\langle r\rangle  \sim 1000$ km, whereas 
the shock of s24.0 (in Figure \ref{fig-entall1}(i)) still wobbles
 around at $\langle r\rangle \sim$ 200 km. This 
demonstrates that the ZAMS mass is not a good 
criterion to diagnose the possibility of explosion.

\begin{figure}
\begin{center}
\FigureFile(90mm,90mm){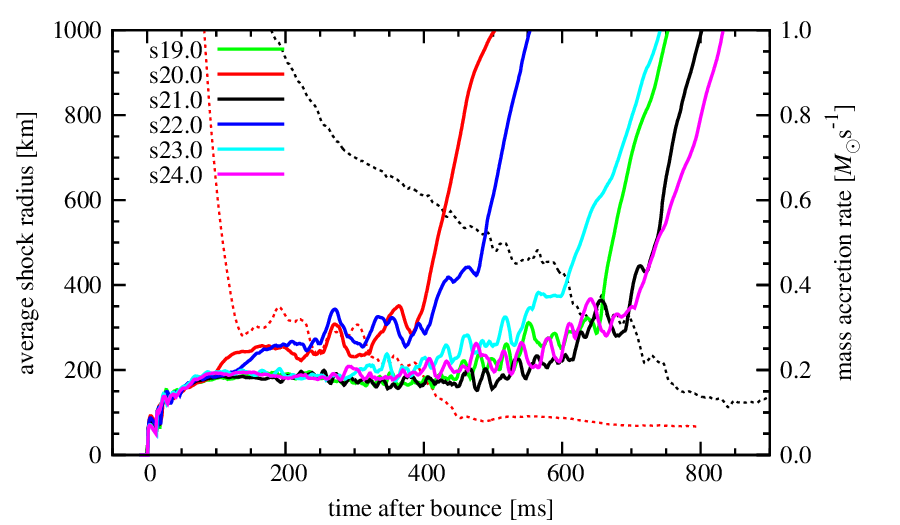}
\end{center}
\caption{Average shock radii (thick solid lines) 
and mass-accretion rate of the collapsing stellar core 
at 500 km (thin dashed lines) 
for some selected models. 
}
\label{fig-rsh}
\end{figure}

This is more clearly visualized 
in Figure \ref{fig-rsh}, showing time evolution of average shock radii
for six models in the mass range 
between $19.0$ $M_{\odot}$ and $24.0 ~M_{\odot}$. 
The shock revival is shown to occur 
earlier for s20.0 (red line) and s22.0 (blue line) compared 
to the lighter progenitors s19.0 (green line) and s21.0 (black line).
Comparing with Figure \ref{fig-xi}, 
it can be seen that the compactness parameter (Equation (\ref{eq-xi}))
is smaller for s20.0 
and s22.0 
in the chosen mass range. 
The smaller compactness is translated into smaller 
mass accretion rate onto the stalled bounce shock. 
For model s20.0 ($\xi_{2.0}=0.19$, red line in Figure \ref{fig-rsh}), 
the relatively earlier shock revival ($\sim 100$ ms postbounce)
coincides with 
the sharp decline of the accretion rate (dashed red line). 
After that, the accretion rate gradually decreases to 
$\sim 0.1 ~M_{\odot} \, {\rm s}^{-1}$ till $t_{400} =420$ ms at this time
 the revived shock has expanded to an average radius of
$\langle r \rangle = 400$ km. 
Here $t_{400}$ is a useful measure to qualify the vigor of the shock revival 
(e.g., \cite{hanke12}). 
On the other hand, model s21.0 has high compactness 
($\xi_{2.0} = 0.46$), 
which leads to the high accretion rate 
(black dashed line in Figure \ref{fig-rsh}). 
It takes 
$\sim$ 500 ms for the sloshing shock of model s21.0 
(black solid line in the Figure \ref{fig-rsh}) 
to gradually turn into a pronounced expansion later on 
and 700 ms to arrive at $\langle r \rangle = 400$ km ($t_{400}$ = 700 ms). 

Note in Figure \ref{fig-rsh} that the correlation between the compactness 
and $t_{400}$ is rather weak. In the Appendix 2, we attempt 
to find the alternative compactness parameter by which the correlation 
is slightly improved.

\begin{figure}
\begin{center}
\FigureFile(90mm,90mm){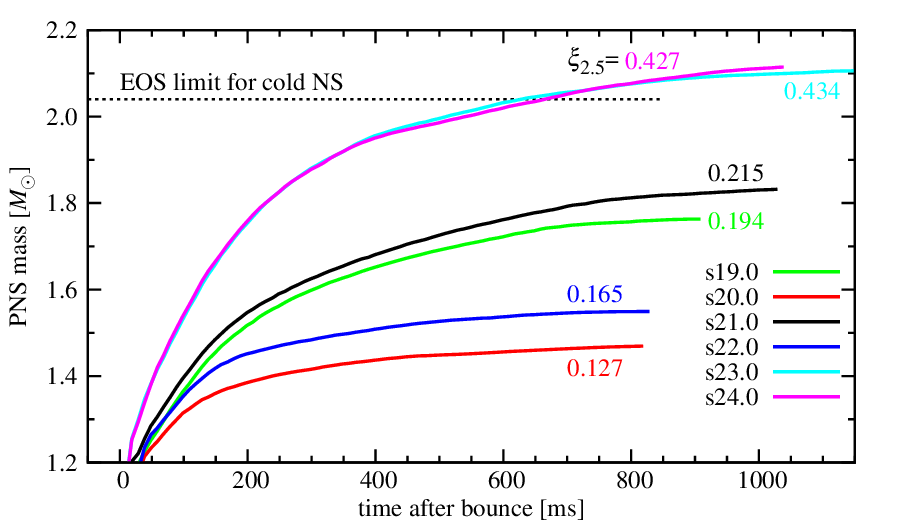}
\end{center}
\caption{Time evolution of central PNS mass 
for the same models as in Figure \ref{fig-rsh}. 
The compactness parameter $\xi_{2.5}$ is labelled beside each line. 
The horizontal dotted line represents the maximum mass 
of a cold neutron star of the LS220 EOS.
}
\label{fig-mpns}
\end{figure}

\begin{figure}
\begin{center}
\FigureFile(90mm,90mm){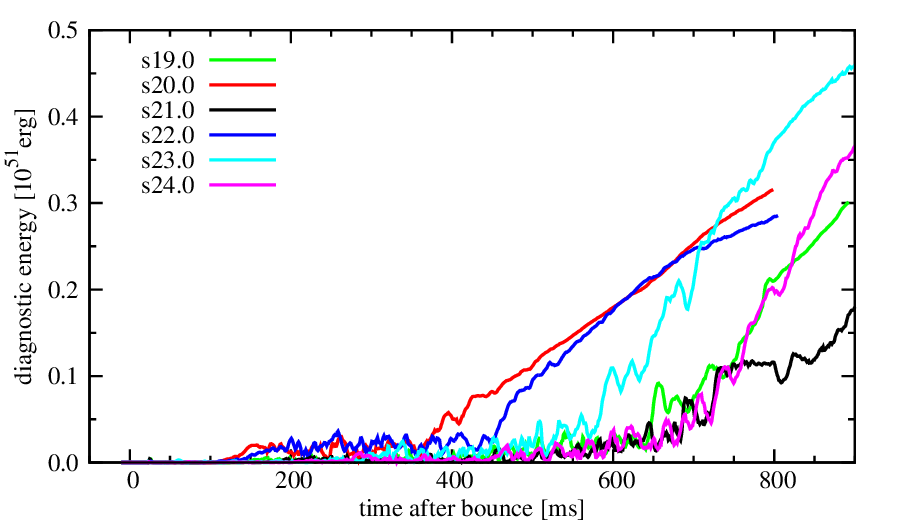}
\end{center}
\caption{Same as Figure \ref{fig-mpns}
but of the diagnostic energy of the explosion.
}
\label{fig-edia}
\end{figure}

The gravitational mass of PNS 
and the {\it diagnostic} explosion energy\footnote{
Following \citet{suwa10} and \citet{nakamura14}, 
we define the {\it diagnostic} energy 
that refers to the integral of the energy over all outward moving zones 
that have a positive sum of the specific internal, kinetic, and 
gravitational energy.} are also shown 
in Figures \ref{fig-mpns} and \ref{fig-edia} 
as a function of post-bounce time. 
Here the PNS is defined by the region 
where the density $\rho > 10^{11}~{\rm g\, cm}^{-3}$. 
The PNS mass is almost converged in our simulation time and 
the value at the final simulation time $t=t_{\rm fin}$ 
has a clear correlation with the compactness parameter. 
In fact, 
the PNS masses become smaller for models with smaller compactness parameter 
(s20.0 drawn in red line and s22.0 in blue) 
 and bigger for models with higher compactness
(s23.0 in sky blue and s24.0 in magenta) 
 and the other two models
(s19.0 in green and s21.0 in black) 
have the intermediate values. 

In Figure \ref{fig-mpns}, the horizontal dotted line represents 
the maximum gravitational mass ($2.04 M_{\odot}$) 
of a cold neutron star (NS) 
for the LS220 EOS employed in this work. 
As seen, the PNS masses of our most ``compact'' models 
(e.g., s23.0 (sky blue line) and s24.0 (magenta line)) exceed the
 limit. Here it should be noted that the above threshold
 is for a cold NS, whereas the PNS soon after bounce is 
still hot. 
At this phase, the contribution of thermal pressure 
to the maximum mass cannot be neglected, 
so that the maximum mass of the hot PNS is bigger 
than that of the cold NS 
\citep{Oconnor11,hempel12}. 
Based on a systematic 1D GR simulation with approximate neutrino transport,
 \citet{Oconnor11} showed that the maximum gravitational 
mass of the hot PNSs, which is bigger for models with high compactness, ranges from $2.1 M_{\odot}$ ($\xi_{2.5,{\rm cb}} = 0.20$)\footnote{Note that 
in \citet{Oconnor11} 
the compactness parameter is estimated at the bounce time 
($\xi_{M,{\rm cb}}$) using LS180 EOS.} 
to $2.5 \, M_\odot$ ($1.15$). From Figure 7 in \citet{Oconnor11}, one can read
 that the maximum gravitational mass for models 
with $\xi_{2.5,{\rm cb}} = 0.4$ (corresponding to the highest $\xi$ in our 
solar-metallicity model series) is $\sim 2.2 \, M_\odot$, 
which is bigger than that of the most massive PNS in our solar metallicity 101 models 
($M_{\rm PNS} = 2.16 \, M_\odot$ for s23.4 model with $\xi_{2.5} = 0.4273$, 
see also Figure \ref{fig-all2.5}). 
In their 1D GR study, 
a model with $\xi_{2.5,{\rm cb}} > 0.4$ 
leads to a BH formation at $t_{\rm pb} \lesssim 1$ s. 
For a given BH-forming progenitor model, 
the BH formation timescale might be delayed in our 2D exploding models 
because the shock expansion would possibly make the mass accretion 
onto the PNS smaller. 
Although multi-D GR simulations with 
elaborate neutrino transport scheme are needed to unambiguously 
clarify this issue, above exploratory discussions lead us to speculate
that the BH formation is less likely to affect the systematic features 
obtained in our solar metallicity models. 
We will comment further on the possible effects of the BH formation 
 in section \ref{sec-uz} including metal-deficient progenitors.

Regarding the diagnostic energy of the explosion (Fig.\ref{fig-edia}), 
it is still increasing at $t=t_{\rm fin}$ 
and its converged value is difficult to be inferred from our simulations. 
The time when the diagnostic energy turns upward corresponds to 
the time of shock revival ($t_{400}$). 
We define its 
increasing rate as
$\dot{E}_{\rm dia} \equiv E_{\rm dia}(t=t_{\rm fin})/(t_{\rm fin} - t_{400})$ 
in units of $10^{51}\,{\rm erg \, s}^{-1}$ 
and find that it 
tends to become higher for models with high compactness 
(0.754, 1.12, and 1.44, for s22.0, s23.0 and s24.0, respectively).
This also indicates that the 
diagnostic energy of the high-$\xi$ models would become higher 
later on.
Note that in previous 1D studies with simplified 
neutrino heating and cooling scheme \citep{Oconnor11} 
or with the excision inside the PNS \citep{Ugliano12},
the relation between the compactness and 
these explosion properties 
could not be determined in a self-consistent manner.

\begin{figure*}
\begin{center}
\FigureFile(170mm,170mm){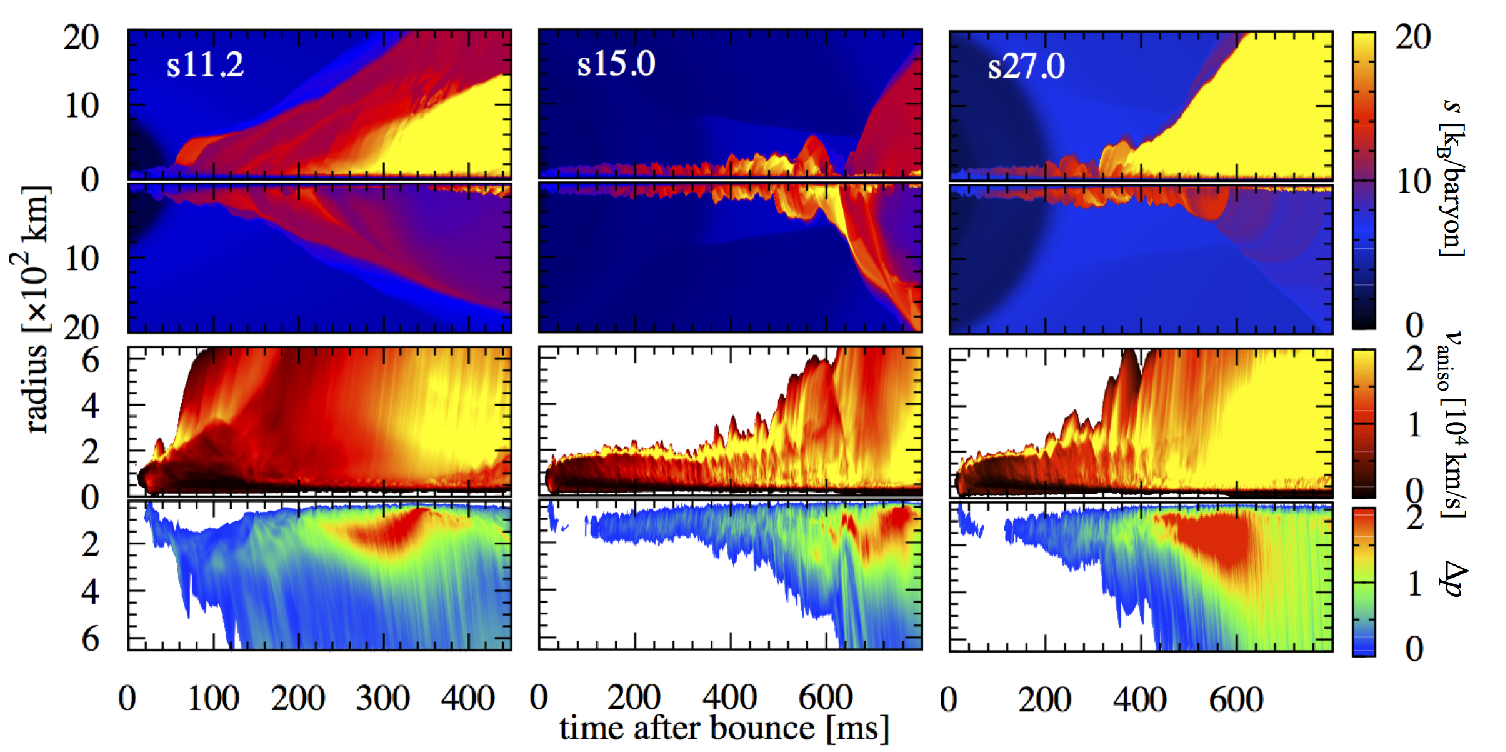}
\end{center}
\caption{
Time-space diagrams of 
specific entropy ($k_{\rm B}$/baryon, $k_{\rm B}$ is the Boltzmann constant), anisotropic velocity (${\rm cm}/{\rm s}$), and pressure perturbation 
for models s11.2 (left), s15.0 (middle), and s27.0 (right). 
Top: the specific entropy 
along the north (upper panels) and south pole (lower). 
Bottom: time evolution of the anisotropic velocity $v_{\rm aniso}$ (upper)
and the normalized pressure perturbation $\Delta p$ (lower) 
in the post-shock region (see the text for details).
}
\label{fig-pol}
\end{figure*}

We investigate the postbounce shock evolution in more details
for three specific models, s11.2, s15.0, and s27.0. 
The density profile of model s11.2 falls rapidly off with radius and 
the compactness parameter 
$\xi_{2.0}$=0.014 is one of the smallest values 
among the 101 progenitors, 
whereas models s15.0 and s27.0 have relatively high compactness 
($\xi_{2.0}=0.298$ and 0.326) 
casting more extended envelope out to the iron core 
(see Figure \ref{fig-mr} for s15.0 model).
Color-coded panels in Figure \ref{fig-pol} show 
the different postbounce evolutions 
for s11.2 (left), s15.0 (middle) and s27.0 (right). 
Model s11.2 explodes rather early 
($t_{400}$ = 150 ms) 
and convective activity as well as the oscillations of the shock is moderate 
before the onset of the explosion 
(see the bottom left panel).
Note in the bottom panels that 
the anisotropic velocity $v_{\rm aniso}$ ({\it upper}), and the 
normalized  pressure perturbation $\Delta p$ ({\it lower}),
is defined respectively as $v_{\rm aniso} = \sqrt{\langle 
\rho [(v_r - \langle v_r \rangle)^2 + v_{\theta}^2]\rangle
/\langle \rho \rangle} $ and $\Delta p = \sqrt{ \langle p^2 \rangle 
- \langle p \rangle^2} /\langle p\rangle$,
where $\langle A \rangle$ represents the angle average of 
quantity $A$ (e.g., \cite{Takiwaki12,kuroda12}). 

For model s15.0 with higher compactness parameter, 
sloshing motions of the shock are 
more clearly visible (middle panels of Figure \ref{fig-pol}). 
These features regarding the dominance of the SASI 
over neutrino-driven convection for high-$\xi$ models 
are quantitatively consistent with those observed in previous 2D simulations 
with more detailed neutrino transport scheme (e.g., \cite{BMuller12}). 
The accreting flows receive an abrupt deceleration 
near at the bottom of the gain region, 
below which the regions are convectively stable. 
A strong pressure perturbation forms there 
(seen, in the lower part of the bottom-middle panel in Figure \ref{fig-pol}, 
as a boundary between the regions colored by white and deep blue at a 
radius of $r \lesssim 100$ km after $t_{\rm pb} \sim 100$ ms). 
In the so-called advective acoustic cycle (e.g.,
 \citet{thierry15} for a review), the pressure perturbations subsequently propagate outward before they hit the shock. 
This leads to the formation of the next vortices (e.g.,
 yellow regions behind the shock in the $v_{\rm aniso}$ plot).
 These features, as previously identified, are natural outcomes of
the SASI and neutrino-driven convection (e.g., \cite{rodrigo14} 
and references therein). 
When the residency timescale becomes enough long 
compared to the neutrino-heating timescale in the gain region 
due to these multi-D effects, 
the runaway shock expansion initiates 
at $t_{\rm pb} \sim$ 500 ms ($t_{400} = 556$ ms) for this model. 

Here we emphasize that the use of the leakage scheme, together
with the omission of inelastic neutrino scattering on electrons 
is likely to facilitate 
artificially easier explosions \citep{taki14}. 
Another caveat is GR effects that cannot be treated in our Newtonian 
simulations. 
Discrepancies between our Newtonian models and GR models 
 might become remarkable 
especially for progenitors with high compactness 
because our simulations show that 
the high-compactness models leave more massive PNSs. 
Comparing model s27.0 ($\xi_{2.5}=0.232$) for example, 
the GR model in \citet{BMuller12} presents 
more rapid revival of the shock ($t_{400} \sim 205$ ms) 
than our Newtonian model ($t_{400} = 432$ ms). 
According to \citet{mueller12b}, 
GR models lead to higher luminosities and mean energies of neutrinos 
and therefore to larger heating efficiencies in the gain layer, 
which is favorable to an explosion. 
For models with the moderate compactness parameters, 
we like to note that some key hydrodynamic features in the 
postbounce phase (such as the onset time of shock 
revival $t_{400}$) are rather similar 
between ours versus the Garching models 
based on \citet{Hanke13} 
($t_{400}$ = 640 vs. 580 ms for model s18.4 ($\xi_{2.5}$=0.188), 
320 vs. 400 ms for s19.6 (0.119), 
540 vs. 560 ms for s21.6 (0.181), 
460 vs. 460 ms for s22.4 (0.200), 
respectively, e.g., Janka et al., TAUP Conference, 2013).  

\begin{figure}
\begin{center}
\FigureFile(90mm,90mm){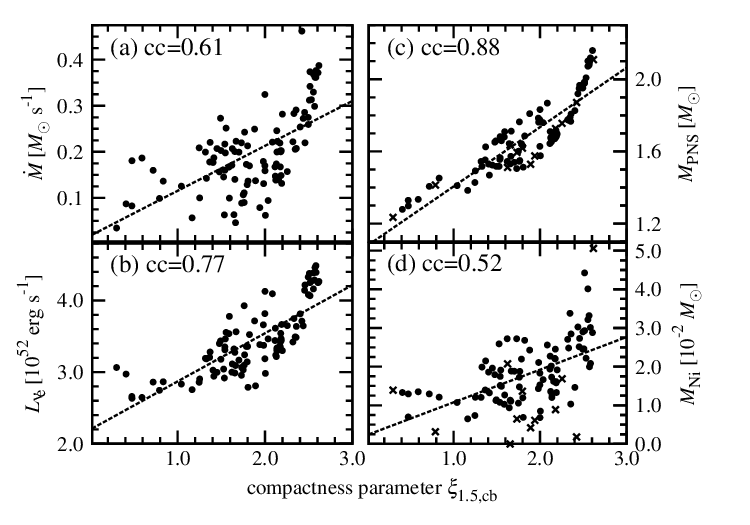}
\end{center}
\caption{
Resultant supernova properties from our 101 simulations 
as a function of the compactness parameter $\xi_{1.5,{\rm cb}}$.
Left: 
(a) Mass accretion rate $\dot{M}$, 
and (b) electron neutrino luminosity $L_{\nu e}$, 
estimated at time of shock revival $t_{400}$.
Right: 
(c) mass of PNS $M_{\rm PNS}$, 
and (d) mass of nickel $M_{\rm Ni}$ in outgoing unbound material, 
at final time of our simulations $t_{\rm fin}$. 
Dashed lines present linear fitting 
with correlation coefficient denoted in each panel. 
In the right panels, 
failed models which cannot carry the shock to the outer boundary 
during our simulation time are shown by cross 
and excluded when we estimate the correlation coefficient.
}
\label{fig-all1.5}
\end{figure}

\begin{figure}
\begin{center}
\FigureFile(90mm,90mm){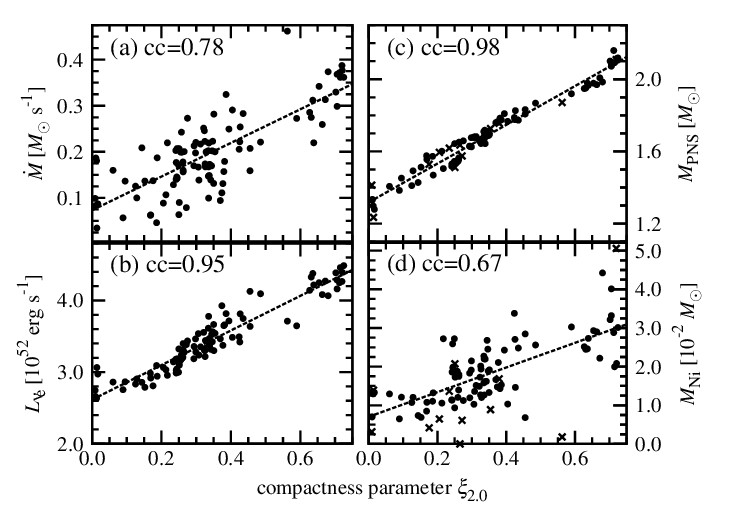}
\end{center}
\caption{
Same as Figure \ref{fig-all1.5} but as a function of $\xi_{2.0}$.
}
\label{fig-all2.0}
\end{figure}

\begin{figure}
\begin{center}
\FigureFile(90mm,90mm){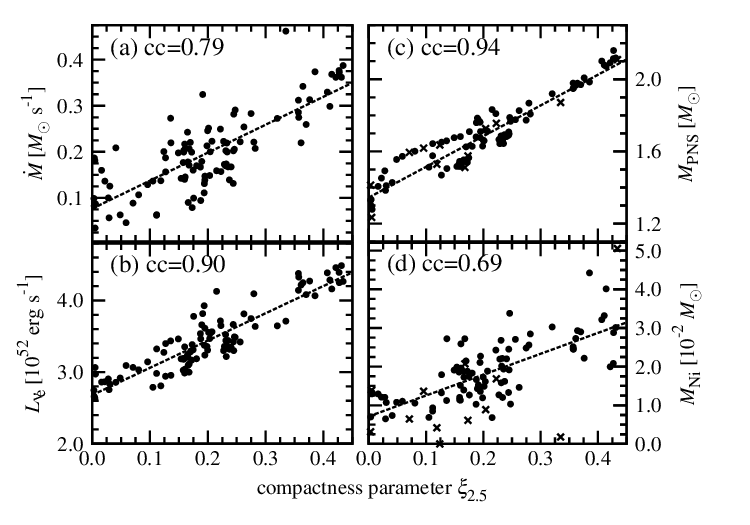}
\end{center}
\caption{
Same as Figure \ref{fig-all1.5} but as a function of $\xi_{2.5}$.
}
\label{fig-all2.5}
\end{figure}

\section{CCSN Properties and Compactness}\label{sec-xi}
In Figures \ref{fig-all1.5} - \ref{fig-all2.5}, 
we plot various quantities to summarize 
the 101 solar-metallicity models as a function of the 
 three different choices of the compactness parameters
 (Fig.12 ($\xi_{1.5, \rm cb}$), Fig.13 ($\xi_{2.0}$), and Fig.14 ($\xi_{2.5}$)). 
In each figure, the left panels show the mass accretion rate $\dot{M}$ and 
electron-type neutrino luminosity $L_{\nu_{\rm e}}$ 
estimated at radius $r=500$ km at time $t=t_{400}$. 
Two panels in the right columns show 
the PNS mass $M_{\rm PNS}$ and the mass of nickel in the ejected material 
$M_{\rm Ni}$ at the final simulation time $t=t_{\rm fin}$.
Each quantity is fitted by a linear line 
and each panel contains a correlation coefficient (cc) defined as
\begin{equation}
cc \equiv \frac{\Sigma_i (\xi_i - \bar{\xi})(y_i-\bar{y})}{\sqrt{\Sigma_i (\xi_i - \bar{\xi})^2} \sqrt{\Sigma_i (y_i - \bar{y})^2}},
\end{equation}
where $\bar{y}$ is a arithmetic mean of quantity $y_i$. 

For most of our models (89 models) $t_{\rm fin}$ is defined as the time 
when the maximum shock radius touches the outer boundary. 
In the remaining 12 models, the shock of model s25.0 has not yet reached
 the outer boundary within our simulation time ($t_{\rm sim} = 1.5$ s).
The other eleven models are stopped during $t_{\rm sim}$  
because the density near the outer boundary of the computation domain 
goes below the lowest value covered by the EOS table. 
Fortunately, the shock revival occurs in these eleven models 
before we stopped the simulations,  
so that we can measure $\dot{M}$ and $L_{\nu_{\rm e}}$ at $t=t_{400}$. 
These incomplete models are not taken into account when we estimate the 
correlation coefficients for $M_{\rm PNS}$ and $M_{\rm Ni}$ at $t=t_{\rm fin}$.
Note that it is technically challenging to extend the EOS table smoothly to 
the non-NSE regime 
where our 13-species alpha network needs to be also consistently treated 
between the two regimes. We leave this for future work.

\citet{Ugliano12} were the first to show that various quantities shown
 in our Figures 12 - 14 are not a monotonic function of the ZAMS mass. 
We confirm it in our 2D simulations 
and find that these values are nearly in the linear 
correlation with the compactness parameters. Furthermore we point out
 that the correlation becomes higher for the choice of $\xi_{2.0}$ 
or $\xi_{2.5}$ compared to $\xi_{1.5, \rm{cb}}$. 
This can be interpreted as follows: 
the core of high-$\xi$ models is surrounded by high-density Si/O layers 
and the mass accretion rate therefore remains high 
long after the stalled shock has formed 
(Figures \ref{fig-all2.0}(a) and \ref{fig-rsh}). 
This makes the PNS mass of the high-$\xi$ models heavier
(Figure \ref{fig-all2.0}(c)). 
Due to the high accretion rate, 
the accretion neutrino luminosities become higher for models 
with high $\xi$ (Figure \ref{fig-all2.0}(b), 
see also \cite{oconnor13}). 
As a result, 
we obtain a stronger shock revival powered by the more intense 
neutrino heating, which makes the amount of the synthesized 
nickel bigger (Figure \ref{fig-all2.0}(d)).


\begin{figure}
\begin{center}
\FigureFile(90mm,90mm){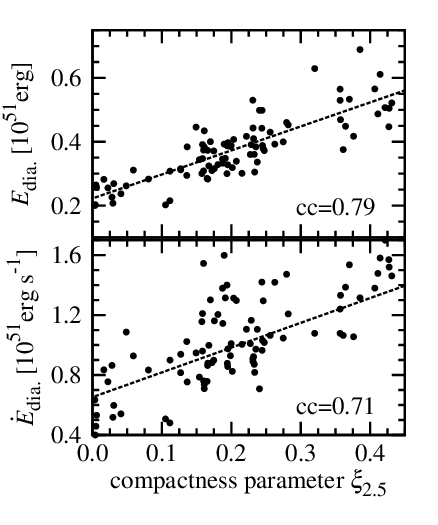}
\end{center}
\caption{
The diagnostic energy of the explosion at $t=t_{\rm fin}$ (top panel) 
and its growth rate (bottom) 
as a function of $\xi_{2.5}$.
}
\label{fig-edia_all}
\end{figure}

The diagnostic energy of the explosion for the 101 models 
is in the range between $\sim 0.2$ - $0.7 \times 10^{51}$ erg, which 
is still increasing at the final time of our simulation.
(see Figure \ref{fig-edia}). 
To obtain a converged value of the diagnostic energy, 
we need to perform a very long-term simulation including the
 special care about the smooth transition of the EOS to the non-NSE regime,
 which is beyond the scope of this work.
Instead of the converged value, 
we estimate the growth rate of the diagnostic energy $\dot{E}_{\rm dia}$ 
defined in Section \ref{sec-res} 
and plot it in Figure \ref{fig-edia_all}. 
Both of the diagnostic energy and its growth rate are 
moderately correlated with the compactness parameter.

\section{Metal-deficient Progenitors}\label{sec-uz}
In this section we move on to discuss the results 
 of metal-deficient progenitors.
The ultra metal-poor models of $10^{-4} ~Z_\odot$ 
contains 247 models with the mass increment of
$0.2 ~M_\odot$ between 11 and $65 ~M_\odot$ 
and a $75 ~M_\odot$ model. The masses of the zero-metallicity progenitors 
are 11 to $40 ~M_\odot$ in every $1.0 ~M_\odot$ (30 models in this series). 

Figure 16 compares $\xi_{2.5}$ of these low metallicity models with 
that of the solar-metallicity models (Figure \ref{fig-xi}). 
Since the metal-deficient models experience no mass loss during
 the stellar evolution,  the compactness of the metal-poor stars
is shown to be much higher for $\gtrsim 30 ~M_\odot$ 
than that of the solar-metallicity models.

\begin{figure}
  \begin{center}
    \FigureFile(90mm,90mm){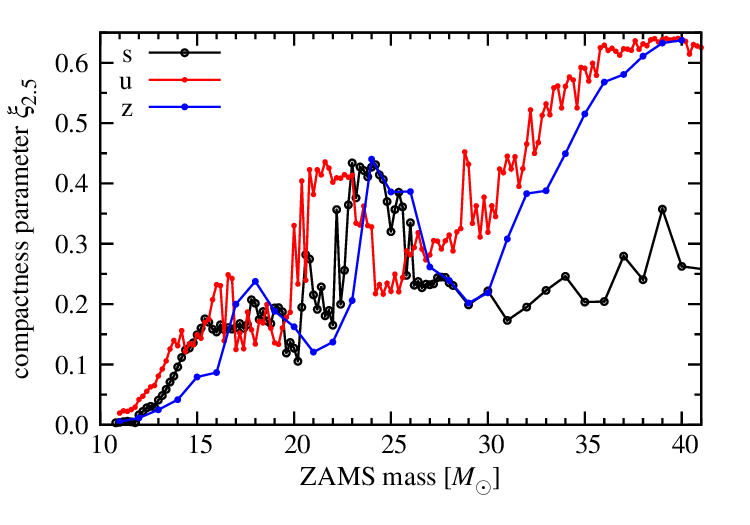}
  \end{center}
  \caption{Compactness parameters $\xi_{2.5}$ of all progenitors as a function of ZAMS mass. 
  Labels s, u, and z, denotes solar-$Z_\odot$, ultra metal-poor $10^{-4}~Z_\odot$, 
  and zero metallicity, respectively. 
  }\label{fig-xi2}
\end{figure}

\begin{figure}
  \begin{center}
    \FigureFile(90mm,90mm){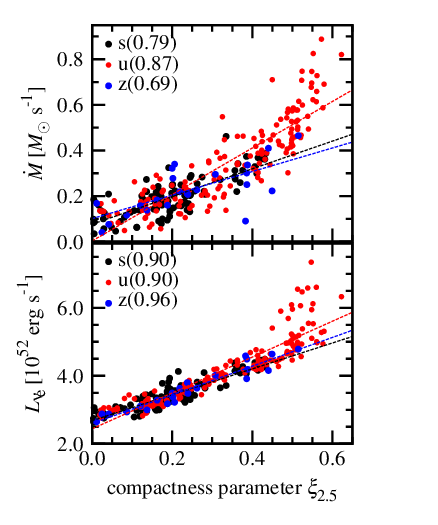}
  \end{center}
  \caption{
  Mass accretion rate $\dot{M}$ (top panel) 
  and electron neutrino luminosity $L_{\nu e}$ (bottom) 
  as a function of the compactness parameter $\xi_{2.5}$.
  These two quantities are estimated 
  at time of shock revival $t_{400}$ for all models.
  Dashed lines present linear fitting 
  with correlation coefficient denoted in each panel for each progenitor type. 
  }\label{fig17}
\end{figure}

Figure \ref{fig17} shows the mass accretion rate $\dot{M}$ (top panel) 
and the electron-neutrino luminosity $L_{\nu e}$ (bottom)  
as a function of the compactness parameter $\xi_{2.5}$.
These two quantities are estimated 
at the time of shock revival $t_{400}$. 
Regardless of the different initial metallicity, the two
 quantities show a similar increasing trend. 
In particular the electron neutrino luminosity is well fitted 
(the correlation coefficients $\geq 0.90$) 
by a linear line. 
Some of the ultra metal-poor progenitors 
with high $\xi_{2.5}$ ($>0.45$, which do not appear in solar-metallicity models)
present very high accretion rate and neutrino luminosity 
above the linear trend. 

\begin{figure}
  \begin{center}
    \FigureFile(90mm,90mm){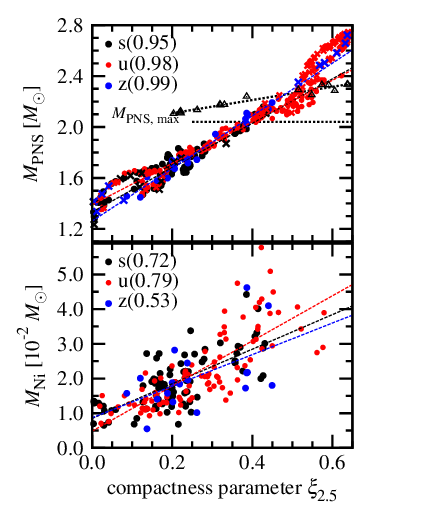}
  \end{center}
  \caption{
  Mass of PNS ($M_{\rm PNS}$, top panel) 
  and mass of nickel in outgoing unbound material ($M_{\rm Ni}$, bottom) 
  as a function of the compactness parameter $\xi_{2.5}$.
  These two quantities are estimated 
  at final time of our simulations ($t = t_{\rm fin}$). 
  Failed models which cannot carry the shock to the outer boundary 
  during our simulation time are shown by crosses 
  and excluded when we estimate the correlation coefficient.
  The horizontal dotted line in the top panel shows 
  the maximum mass of cold NS for LS220 EOS ($2.04 \, M_\odot$).
  The another dotted line in the top panel is 
  a linear fit of the triangles, the maximum PNS masses 
  of BH forming models in \citet{Oconnor11} 
  (see the text for detail).
 }\label{fig18}
\end{figure}

Figure \ref{fig18} shows the PNS mass $M_{\rm PNS}$ (top panel) 
and the mass of the synthesized nickel $M_{\rm Ni}$ in the 
outgoing unbound material (bottom) 
as a function of the compactness parameter $\xi_{2.5}$. 
These two quantities are estimated 
at final time of our simulations $t_{\rm fin}$. 
All of the models again show a similar increasing trend. 
Crosses in the top panel of Figure \ref{fig18} 
represents the models in which the revived shock did not reach the outer 
boundary during our simulation time ($t = $1.5 s)\footnote{ 
Some of them lying in low $\xi_{2.5}$ ($\lesssim 0.4$) 
are caused by a numerical reason, that is, as we have already mentioned,
our simulations are stopped 
when a low density region out of our EOS table emerged.}. 
We exclude all the unsuccessful (non-exploding) models 
shown by the crosses in the top panel of Figure \ref{fig18}
when we estimate the correlation coefficients. 

The central PNS of the models in the upper-right corner of Figure \ref{fig18}
would finally collapse to form a black hole (BH), 
although our Newtonian code cannot follow such dynamical behaviors. 
Triangles in the top right panel of Figure \ref{fig18} 
represent the maximum PNS masses 
estimated by \citet{Oconnor11}
in which the same progenitor models \citep{Woosley02} 
and the same EOS (LS220) 
as in this study were examined by their 1D GR code. 
The dotted curve ($M_{\rm PNS, max} = 0.52 \, \xi_{2.5} + 2.01$) 
is a linear fitting to their results. 
Our models above the dividing line should be predominantly 
affected by general relativity and such models cannot be treated 
appropriately by our Newtonian code. 

The number fraction of the progenitor models 
which leave PNS more massive than $M_{\rm PNS, max}$ 
is $104 / 378 \sim 28$ \%. 
Most of them belong to the u-series ($98 / 104$), 
the ZAMS mass of which is bigger than $32.2 M_\odot$. 
Adopting the Salpeter initial mass function to weight the ZAMS mass gives 
$\int_{32.2 M_\odot}^{75 M_\odot} m^{-2.35} dm 
/ \int_{10 M_\odot}^{75 M_\odot} m^{-2.35} dm 
\sim 15$ \%. 
Although the majority of the PNSs in our 2D models is 
less than the maximum mass of PNS, we need to perform 
GR simulations (e.g., \citet{Oconnor11,BMuller12,kuroda12}) in order to
elucidate the fate of the 
high-$\xi$ metal-poor stars.

\section{Conclusions and Discussion}\label{sec-con}
We have presented an overview of 2D core-collapse supernova simulations 
employing neutrino transport scheme by the IDSA scheme.
We studied 101 solar-metallicity, 
247 ultra metal-poor, and 30 zero-metal progenitors covering 
zero-age main sequence mass 
from $10.8 ~M_{\odot} $ to $75.0 ~M_{\odot} $. 
Using the 378 progenitors in total, 
we systematically investigated how the differences in the structures of 
these multiple 
progenitors impact the hydrodynamics evolution.
By following a long-term evolution over 1.0 s after bounce, 
most of the computed models exhibited neutrino-driven revival of the 
stalled bounce shock at $\sim$ 200 - 800 ms postbounce, 
leading to the possibility of explosion. 
Pushing the boundaries of expectations in previous one-dimensional (1D) 
studies, our results confirmed that 
the compactness parameter $\xi$ that characterizes the structure 
of the progenitors is also a key in 2D to diagnose the properties of 
 neutrino-driven explosions.
Models with high $\xi$ 
undergo high ram pressure from the accreting matter onto the stalled shock,
 which affects the subsequent evolution of the shock expansion 
and the mass of the protoneutron star under the influence of 
 neutrino-driven convection and 
the standing accretion-shock instability. 
We have shown that the accretion luminosity becomes higher for models 
with high $\xi$, which makes the growth rate of diagnostic 
energy higher and the synthesized nickel mass bigger. 
We have found that these explosion characteristics 
tend to show a monotonic increase as a function of the 
compactness parameters $\xi$. 

Our simulations are limited in space ($r < 5000$ km) and time ($t \leq 1.5$ s). 
The simulations are terminated before the diagnostic energies are 
saturated. Later on neutrino energy deposition would get smaller with time
as the neutrino luminosity as well as post-shock density becomes smaller. 
Further global simulation, 
taking account of gravitational energy of an envelope and 
nuclear energy released via recombination process behind the shock, 
is necessary to determine the final explosion energy (Figure \ref{fig-edia_all}).
Moreover, the finding of this study should be reexamined 
by 3D models (\cite{hanke12,dolence12,couch13,couch14,nakamura14_2}), 
in which neutrino transport is appropriately solved 
(see, discussions in \cite{Hanke13,taki14,nagakura14,tony14}).
It is also important to study the impacts of 
the precollapse non-spherical structures (e.g., \cite{Arnett11}) 
 on fostering the shock revival (e.g., \cite{couchott}).
To get a more accurate amount of the synthesized nickel and other isotopes, 
a post-process calculation with a larger nuclear network is 
needed. In the more long run, wide-range long-term 3D full-scale GR 
simulations are needed to unambiguously clarify the critical 
 $\xi$ parameter, below or above which neutron stars or black holes 
will be left behind. 

In this work we have reported our results 
of only 
progenitors from one modelling group. 
Currently we are conducting the same sort of CCSN simulations 
for sets of progenitors from the other groups including 
a variety of metallicity, rotation, and magnetic fields, 
which will be presented in the forthcoming work.

We thank F.K Thielemann, M. Liebend\"{o}rfer, R. M. Cabez\'on, 
M. Hempel, and  K.-C. Pan for stimulating discussions and for their kind hospitality during our research stay at Basel in February, 2014.
We are also grateful to S. Yamada and Y. Suwa for informative discussions. 
Numerical computations were carried out in part on 
XC30 and general common use computer system at the center for 
Computational Astrophysics, CfCA, 
the National Astronomical Observatory of Japan,
Oakleaf FX10 at Supercomputing Division in University of Tokyo, and on 
SR16000 at YITP in Kyoto University.
 This study was supported in part by the Grants-in-Aid for the Scientific 
Research from the Ministry of Education, Science and Culture of Japan 
(Nos. 23540323, 23340069, 24103006, 24244036, and 26707013) and by HPCI Strategic Program of Japanese MEXT.

\newpage


\appendix
\section{Numerical Resolution}\label{sec-app}

Numerical resolution should be taken as high as possible in order
to capture accurately hydrodynamics processes in computational
 fluid dynamics.
In the state-of-the-art CCSN simulations, 
especially in multi-D models, however, 
high numerical resolution drastically increases the numerical cost and 
a full convergence has not yet been obtained 
even in simplified simulations. 

\citet{hanke12} investigated the resolution dependence 
 in their 2D and 3D models using 11.2 and 15 $M_\odot$ progenitors 
with a parameterized neutrino heating and cooling scheme. 
In most of the 2D models, better angular resolution led to 
 easier explosion, although some of them do not obey this trend. 
On the other hand, 
\citet{couch13} investigated a $15 M_\odot$ progenitor 
employing the same simple neutrino scheme as in \citet{hanke12}, 
and found that 2D explosions are delayed for models with 
 higher numerical resolution.
More recently, 
\citet{taki14} computed 2D and 3D models for an $11.2 M_\odot$ progenitor 
with an energy-dependent neutrino transport scheme 
and concluded that 
higher numerical resolutions led to slower onset of the shock revival 
in both 2D and 3D. 

In our fiducial 2D runs, the angular resolution ($\Delta \theta$) is taken as $1.4^\circ$
(128 angular zones to cover $0 \leq \theta \leq \pi$). 
In this Appendix, we briefly discuss the resolution dependence, 
 in which we compare the results with the fiducial resolution with 
those with high resolution ($\Delta \theta = 0.7^\circ$ with 
 256 angular zones). 

\begin{figure}
  \begin{center}
    \FigureFile(90mm,90mm){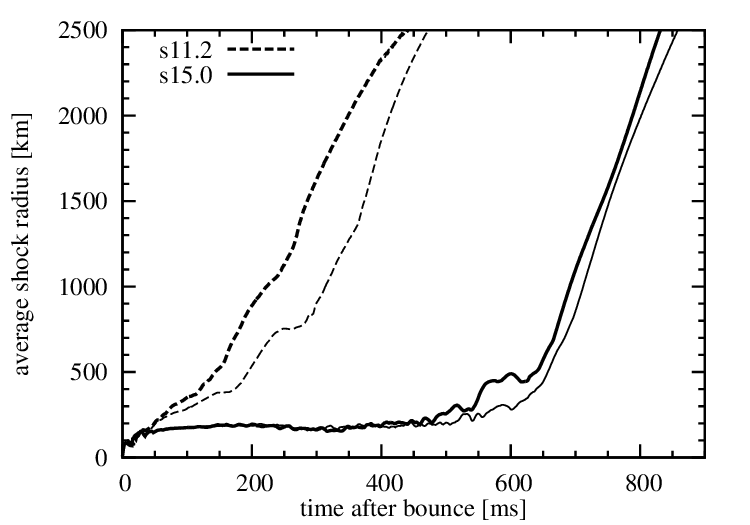}
  \end{center}
  \caption{
Comparison between different resolution models. 
Average shock radii as a function of time relative to bounce 
for s11.2 and s15.0 models are shown. 
The shock revives more rapidly 
in the fiducial resolution models (thick lines) 
than in high resolution models (thin lines).
 }\label{fig-a1}
\end{figure}

Figure \ref{fig-a1} shows time evolution of the average shock radii 
for two progenitors, s11.2 and s15.0. 
In both of the two, it is shown that the shock revival is 
delayed for the high-resolution models (thin lines).
The difference of $t_{400}$, $t_{1000}$, and $t_{2500}$
(the time when the average shock radius arrives 
at 400, 1,000, and 2,500 km, respectively) 
are 30.4 \% (13.1 \%), 29.0 \% (2.9 \%), and 6.8 \% (3.1 \%)
for model s11.2 (s15.0). 
The relatively large difference of $t_{400}$ may reflect 
 that the shock revival time is more likely to be affected by 
 stochastic matter motions driven by neutrino-driven convection and
 the SASI.

\begin{figure}
  \begin{center}
    \FigureFile(90mm,90mm){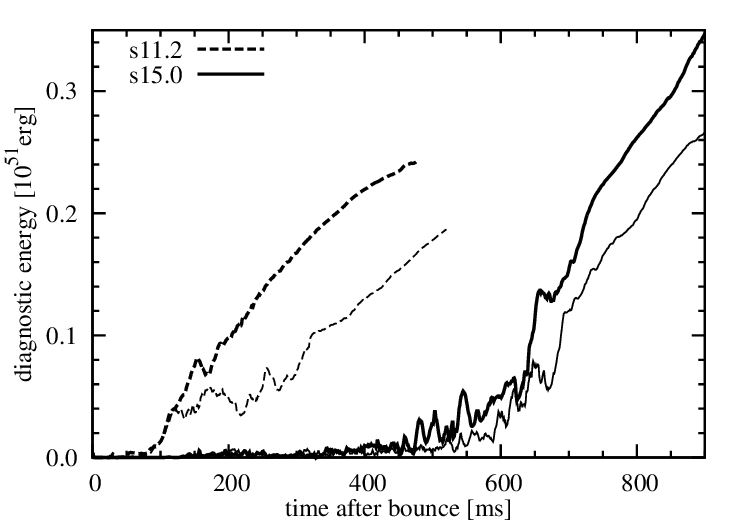}
  \end{center}
  \caption{
Same as Figure \ref{fig-a1} but for the evolution of the 
diagnostic energies of the explosion. 
 }\label{fig-a2}
\end{figure}

The evolution of the diagnostic energies are presented in Figure \ref{fig-a2}. 
Comparing them at $t=t_{1000}$ and $t_{2500}$, 
the differences between the fiducial and high resolution models 
are 22.7 \% (9.8 \%) and 28.9 \% (13.9 \%)
for s11.2 (s15.0) model. 
The difference is more bigger in model s11.2, 
which has a small compactness parameter and is weakly exploding, 
than in model s15.0. 

Our fiducial resolution does not give well-converged results 
at least for model s11.2. 
The systematic behaviors that we found, 
as shown in Figure \ref{fig-all2.0} for example, 
might be subject to change quantitatively.
At least, our 2D results showing that higher numerical resolutions lead to 
slower evolution of the shock radius and the diagnostic explosion energy, 
 are consistent with \citet{couch13} and \citet{taki14}.
A systematic resolution study including a detailed comparison
 between different numerical codes, schemes, and setups should be
 done urgently, which we leave for future work.

\section{Time of Shock Revival}\label{sec-t400}

\begin{figure}
\begin{center}
\FigureFile(90mm,90mm){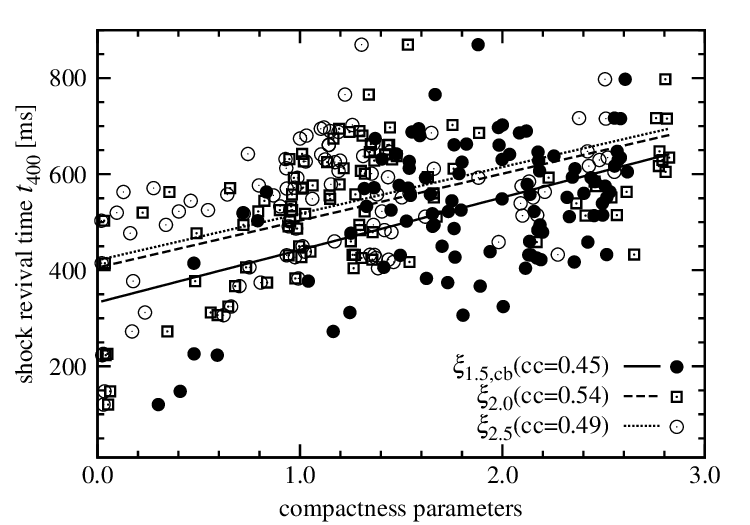} 
\end{center}
\caption{
Time of shock revival $t_{400}$ as three kinds of the compactness parameters. 
$\xi_{2.0}$ and $\xi_{2.5}$ are calibrated by arbitrary factors.
}
\label{fig-t400-1}
\end{figure}

Our systematic 2D CCSN simulations demonstrate that the $\xi$ parameter
 is a good diagnostics to infer the progenitor-remnant and 
progenitor-explosion connections. However, the time of shock revival ($t_{400}$) 
shows a large scatter and weaker correlation with the compactness parameter. 
Figure \ref{fig-t400-1} shows $t_{400}$ 
for three kinds of the compactness parameters 
and we obtain low correlation coefficients (0.45-0.54). 
This may partly come from 
the stochasticity of the nonlinear growth of SASI and convection, 
seeded by initial random perturbations, 
which affects the subsequent shock evolution. 
Another possibility is that our definition of the compactness parameters
might be inappropriate to characterize $t_{400}$. 
In this Appendix, 
we attempt to find a more appropriate form of the compactness parameter 
to characterize the shock revival time. 

As we have seen in Figure \ref{fig-rsh}, 
at least for some models,
the shock revival time seems to be linked 
to the time when the mass accretion rate drops. 
It typically occurs when the Si/O interface falls onto the shock. 
To capture this, 
 it would be better to estimate the compactness 
at the Si/O interface which differs from models to models. 
In fact, models with earlier shock revival tend to have a more compact
 Si layer. 
Here we define an another compactness parameter $\xi_{\rm i}$ as
\begin{equation}
\xi_{\rm i} \equiv 
\frac{\Delta M_{\rm i}/M_{\odot}}{\Delta R_{\rm i}/1000{\rm km}},
\end{equation}\label{eq-xi2}
where subscript $i$ denotes representative element at a certain layer, 
 $\Delta R_{\rm i}$ and $\Delta M_{\rm i}$ corresponds to
the width and included mass in the layer, respectively.

\begin{figure}
\begin{center}
\FigureFile(90mm,90mm){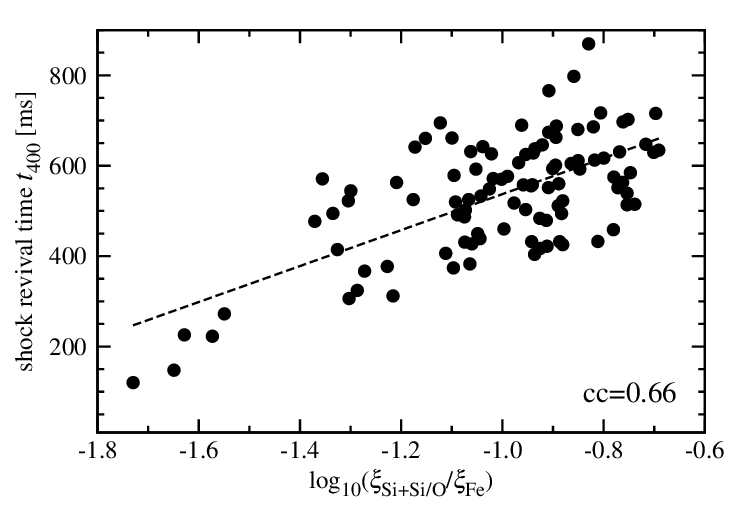} 
\end{center}
\caption{
Time of shock revival $t_{400}$ estimated as a function of the 
 alternative compactness parameter (e.g., Eq. (A1)).
}
\label{fig-t400-2}
\end{figure}

Figure \ref{fig-t400-2} presents $t_{400}$ 
as a function of the ratio $\xi_{\rm Si+Si/O}/\xi_{\rm Fe}$. 
The concept of this alternative parameter is as follows. 
The compactness parameter defined at the surface of iron core 
($\xi_{\rm Fe}$) is a measure to predominantly determine 
the core neutrino luminosity, whereas $\xi_{\rm Si+Si/O}$ is a measure to
 the density decline in the outer layer. So we expect that the smaller 
 ratio would lead to an easier explodability.
As shown in Figure 22, this alternative indicator 
gives the correlation coefficient of 0.66, which is better than 0.49
 estimated from $\xi_{2.5}$. To enhance the predicative power, we 
should more carefully analyze how the compactness parameters are related 
 to the core/accretion luminosity, and the density jump at 
Si/O interface. We leave this for the future work. 

\end{document}